\documentclass[aps,preprint,preprintnumbers,amsmath,amssymb,nofootinbib]{revtex4}
\usepackage{amssymb}
\usepackage{lscape}
\usepackage{amsmath}
\usepackage{bm}
\usepackage{graphicx}
\usepackage{epsfig}
\begin{document}
\title{Charged and electromagnetic fields from relativistic quantum geometry}
\author{$^{2}$ Marcos R. A. Arcod\'{\i}a\footnote{E-mail address: marcodia@mdp.edu.ar},  $^{1,2}$ Mauricio Bellini
\footnote{E-mail address: mbellini@mdp.edu.ar} }
\address{$^1$ Departamento de F\'isica, Facultad de Ciencias Exactas y
Naturales, Universidad Nacional de Mar del Plata, Funes 3350, C.P.
7600, Mar del Plata, Argentina.\\
$^2$ Instituto de Investigaciones F\'{\i}sicas de Mar del Plata (IFIMAR), \\
Consejo Nacional de Investigaciones Cient\'ificas y T\'ecnicas
(CONICET), Mar del Plata, Argentina.}

\begin{abstract}
In the Relativistic Quantum Geometry (RQG) formalism recently introduced, was explored the possibility that the
variation of the tensor metric can be done in a Weylian
integrable manifold using a geometric displacement, from a Riemannian to a Weylian
integrable manifold, described by the dynamics of an auxiliary geometrical scalar field $\theta$, in order that the
Einstein tensor (and the Einstein equations) can be represented on
a Weyl-like manifold.
In this framework we study jointly the dynamics of electromagnetic fields produced by quantum complex vector fields, which describes charges without charges. We demonstrate that complex fields act as a source of tetra-vector fields which describe an
extended Maxwell dynamics.
\end{abstract}
\maketitle

\section{Introduction}

The consequences of non-trivial topology for the laws of physics has been a topic of perennial interest for theoretical physicists\cite{Weyl}, with applications to non-trivial spatial topologies\cite{cve} like Einstein-Rosen bridges, wormholes, non-orientable spacetimes, and quantum-mechanical entanglements.

Geometrodynamics\cite{Wh} is a picture of general relativity that study the evolution of the spacetime geometry. The key notion of the Geometrodynamics was the idea of {\it charge without charge}. The Maxwell field  was taken to be source free, and so a non-vanishing charge could only arise from an electric flux lines trapped in the topology of spacetime. With the construction of ungauged supergravity theories it was realised that the Abelian gauge fields in such theories were source-free, and so the charges arising therein were therefore central charges\cite{G1} and as consequence satisfied a BPS bound\cite{G2} where the embedding of Einstein-Maxwell theory into $N=2$ supergravity theory was used. The significant advantages of geometrodynamics, usually come at the expense of manifest local Lorentz symmetry\cite{R}. During the 70's and 80's decades a method of quantization was developed in order to deal with some unresolved problems of quantum field theory in curved spacetimes\cite{pru}.

In a previous work\cite{RB} was explored the possibility that the
variation of the tensor metric must be done in a Weylian
integrable manifold using a geometric displacement, from a Riemannian to a Weylian
integrable manifold, described by the dynamics of an auxiliary geometrical scalar field $\theta$, in order that the
Einstein tensor (and the Einstein equations) can be represented on
a Weyl-like manifold. An important fact is that the Einstein tensor complies with the
gauge-invariant transformations studied in a previous work\cite{rb}. This method is very useful because can be used to describe, for instance, nonperturbative back-reaction effects
during inflation\cite{be2016}. Furthermore, it was introduced the relativistic
quantum dynamics of $\theta$ by using the fact that the cosmological constant $\Lambda$ is a
relativistic invariant. In this letter, we extend our study to complex charged fields that act as the source of vector fields $A^{\mu}$.

\section{RQG revisited}

The first variation of the Einstein-Hilbert (EH) action ${\cal I}$\footnote{Here, $g$ is the determinant of the covariant background tensor metric $g_{\mu\nu}$, $R=g^{\mu\nu} R_{\mu\nu}$ is the scalar curvature,
$R^{\alpha}_{\mu\nu\alpha}=R_{\mu\nu}$ is
the covariant Ricci tensor and ${\cal L}_m$ is an arbitrary Lagrangian density which describes matter. If we deal with an orthogonal base,
the curvature tensor will be written in terms of the connections:
$R^{\alpha}_{\,\,\,\beta\gamma\delta} = \Gamma^{\alpha}_{\,\,\,\beta\delta,\gamma} -  \Gamma^{\alpha}_{\,\,\,\beta\gamma,\delta}
+ \Gamma^{\epsilon}_{\,\,\,\beta\delta} \Gamma^{\alpha}_{\,\,\,\epsilon\gamma} - \Gamma^{\epsilon}_{\,\,\,\beta\gamma}
\Gamma^{\alpha}_{\,\,\,\epsilon\delta}$.}
\begin{equation}\label{act}
{\cal I} =\int_V d^4x \,\sqrt{-g} \left[ \frac{R}{2\kappa} + {\cal L}_m\right],
\end{equation}
is given by
\begin{equation}\label{delta}
\delta {\cal I} = \int d^4 x \sqrt{-g} \left[ \delta g^{\alpha\beta} \left( G_{\alpha\beta} + \kappa T_{\alpha\beta}\right)
+ g^{\alpha\beta} \delta R_{\alpha\beta} \right],
\end{equation}
where $\kappa = 8 \pi G$, $G$ is the gravitational constant and $g^{\alpha\beta} \delta R_{\alpha\beta} =\nabla_{\alpha}
\delta W^{\alpha}$, where  $\delta W^{\alpha}=\delta
\Gamma^{\alpha}_{\beta\gamma} g^{\beta\gamma}-
\delta\Gamma^{\epsilon}_{\beta\epsilon}
g^{\beta\alpha}=g^{\beta\gamma} \nabla^{\alpha}
\delta\Psi_{\beta\gamma}$. When the flux of $\delta W^{\alpha}$ that cross the Gaussian-like hypersurface defined in an arbitrary region of the spacetime, is nonzero, one obtains in the last term of (\ref{delta}), that
$\nabla_{\alpha} \delta W^{\alpha}=\delta\Phi(x^{\alpha})$, such that $\delta\Phi(x^{\alpha})$ is an arbitrary scalar field that takes into account the flux of $\delta W^{\alpha}$
across the Gaussian-like hypersurface. This flux becomes zero when there are no sources inside this hypersurface. Hence, in order to make
$\delta {\cal I}=0$ in (\ref{delta}), we must consider the condition: $
G_{\alpha\beta} + \kappa T_{\alpha\beta} = \Lambda\,
g_{\alpha\beta}$, where $\Lambda$ is the cosmological constant. Additionally, we must require
the constriction $\delta g_{\alpha\beta} \Lambda =
\delta\Phi\, g_{\alpha\beta}$. Then, we propose the existence of a tensor
field $\delta\Psi_{\alpha\beta}$, such that $\delta
R_{\alpha\beta}\equiv \nabla_{\beta} \delta W_{\alpha}-\delta\Phi
\,g_{\alpha\beta} \equiv \Box \delta\Psi_{\alpha\beta} -\delta\Phi
\,g_{\alpha\beta} =- \kappa \,\delta S_{\alpha\beta}$\footnote{We
have introduced the tensor $S_{\alpha\beta} = T_{\alpha\beta}
-\frac{1}{2} T \, g_{\alpha\beta}$, which takes into account
matter as a source of the Ricci tensor $R_{\alpha\beta}$.}, and
hence $\delta W^{\alpha} = g^{\beta\gamma} \nabla^{\alpha}
\delta\Psi_{\beta\gamma}$, with $\nabla^{\alpha}
\delta\Psi_{\beta\gamma}=\delta\Gamma^{\alpha}_{\beta\gamma} -
\delta^{\alpha}_{\gamma} \delta\Gamma^{\epsilon}_{\beta\epsilon}$.
{\em Notice that the fields $\bar{ \delta W}_{\alpha}$ and
$\bar{\delta\Psi}_{\alpha\beta}$ are gauge-invariant under
transformations}:
\begin{equation}
\bar{\delta W}_{\alpha} = \delta W_{\alpha} - \nabla_{\alpha} \delta\Phi, \qquad
\bar{\delta\Psi}_{\alpha\beta} =\delta\Psi_{\alpha\beta} - \delta\Phi \,
g_{\alpha\beta}, \label{gauge}
\end{equation}
where the scalar field $\delta\Phi$ complies $\Box \delta\Phi =0$. On the other hand, we can make the transformation
\begin{equation}\label{ein}
\bar{G}_{\alpha\beta} = {G}_{\alpha\beta} - \Lambda\, g_{\alpha\beta},
\end{equation}
and the transformed Einstein equations with the equation of motion for the transformed gravitational waves, hold
\begin{eqnarray}
&& \bar{G}_{\alpha\beta} = - \kappa\, {T}_{\alpha\beta}, \label{e1} \\
&& \Box \bar{\delta\Psi}_{\alpha\beta} =- \kappa \,\delta
S_{\alpha\beta}, \label{e2}
\end{eqnarray}
with $\Box \delta\Phi(x^{\alpha})=0$ and $\delta\Phi(x^{\alpha})\,
g_{\alpha\beta} =  \Lambda\,\delta g_{\alpha\beta}$. The eq.
(\ref{e1}) provides us the Einstein equations with cosmological
constant included, and (\ref{e2}) describes the exact equation of
motion for gravitational waves with an arbitrary source $\delta
S_{\alpha\beta}$ on a closed and curved space-time. A very
important fact is that the scalar field $\delta\Phi(x^{\alpha})$ appears
as a scalar flux of the tetra-vector with components $\delta W^{\alpha}$
through the closed hypersurface $\partial{\cal M}$. This arbitrary
hypersurface encloses the manifold by down and must be viewed as a
3D Gaussian-like hipersurface situated in any region of
space-time. This scalar flux is a gravitodynamic potential related
to the gauge-invariance of $\delta W^{\alpha}$ and the gravitational
waves $\bar{\delta\Psi}_{\alpha\beta}$. Other important fact is
that since $\delta \Phi(x^{\alpha})\, g_{\alpha\beta} = \Lambda\,\delta
g_{\alpha\beta}$, the existence of the Hubble horizon is related
to the existence of the Gaussian-like hypersurface. The variation of the
metric tensor must be done in a Weylian
integrable manifold\cite{RB} using an
auxiliary geometrical scalar field $\theta$, in order to the
Einstein tensor (and the Einstein equations) can be represented on
a Weyl-like manifold, in agreement with the gauge-invariant
transformations (\ref{gauge}). If we consider a zero covariant derivative of the metric tensor in the Riemannian manifold
(we denote with $";"$ the Riemannian-covariant derivative): $\Delta g_{\alpha\beta}=g_{\alpha\beta;\gamma} \,dx^{\gamma}=0$,
hence the Weylian covariant derivative $ g_{\alpha\beta|\gamma} = \theta_{\gamma}\,g_{\alpha\beta}$, described with respect to the Weylian connections \footnote{To simplify the notation we shall denote $\theta_{\alpha} \equiv \theta_{,\alpha}$}
\begin{equation}\label{ga}
\Gamma^{\alpha}_{\beta\gamma} = \left\{ \begin{array}{cc}  \alpha \, \\ \beta \, \gamma  \end{array} \right\}+ g_{\beta\gamma} \theta^{\alpha},
\end{equation}
will be nonzero
\begin{equation}\label{gab}
\delta g_{\alpha\beta} = g_{\alpha\beta|\gamma} \,dx^{\gamma} = -\left[\theta_{\beta} g_{\alpha\gamma} +\theta_{\alpha} g_{\beta\gamma}
\right]\,dx^{\gamma}.
\end{equation}

\subsection{Gauge-invariance and quantum dynamics}

From the action's point of view, the scalar field $\theta(x^{\alpha})$ is a generic geometrical transformation that leads invariant the action
\begin{equation}\label{aac}
{\cal I} = \int d^4 x\, \sqrt{-\hat{g}}\, \left[\frac{\hat{R}}{2\kappa} + \hat{{\cal L}}\right] = \int d^4 x\, \left[\sqrt{-\hat{g}} e^{-2\theta}\right]\,
\left\{\left[\frac{\hat{R}}{2\kappa} + \hat{{\cal L}}\right]\,e^{2\theta}\right\},
\end{equation}
where we shall denote with a hat, $\, \hat{}\,$, the quantities represented on the Riemannian manifold. Hence, Weylian quantities will be varied over these
quantities in a Riemannian manifold so that the dynamics of the system preserves the action: $\delta {\cal I} =0$, and we obtain
\begin{equation}
-\frac{\delta V}{V} = \frac{\delta \left[\frac{\hat{R}}{2\kappa} + \hat{{\cal L}}\right]}{\left[\frac{\hat{R}}{2\kappa} + \hat{{\cal L}}\right]}
= 2 \,\delta\theta,
\end{equation}
where $\delta\theta = -\theta_{\mu} dx^{\mu}$ is an exact differential and $V=\sqrt{-\hat{ g}}$ is the volume of the Riemannian manifold. Of course, all the variations are in the Weylian geometrical representation, and assure us gauge invariance because $\delta {\cal I} =0$.
Using the fact that the tetra-length is given by $S=\frac{1}{2} x_{\nu} \hat U^{\nu}$ and the Weylian velocities are given by $u^{\mu} = \hat U^{\mu} + \theta^{\mu} \left(x_{\epsilon} \hat U^{\epsilon}\right)$, can be demonstrated that
\begin{equation}
u^{\mu} u_{\mu} = 1 + 4 S \left( \theta_{\mu} \hat U^{\mu} - \frac{4}{3} \Lambda\, S\right).
\end{equation}
The components $u^{\mu}$ are the relativistic quantum velocities, given by the geodesic equations
\begin{equation}
\frac{du^{\mu}}{dS} + \Gamma^{\mu}_{\alpha\beta} u^{\alpha} u^{\beta} =0,
\end{equation}
such that the Weylian connections $\Gamma^{\mu}_{\alpha\beta}$ are described by (\ref{con}). In other words, the quantum velocities
$u^{\mu}$ are transported with parallelism on the Weylian manifold, meanwhile $\hat{U}^{\mu}$ are transported with parallelism on the Riemann manifold.
If we require that $u^{\mu} u_{\mu} = 1$, we obtain the gauge
\begin{equation}\label{gau}
\hat\nabla_{\mu} A^{\mu} = \frac{2}{3} \Lambda^2 \, S(x^{\mu}).
\end{equation}
Hence, we obtain the important result
\begin{equation}
d\Phi = \frac{1}{6} \Lambda^2 \, S\, dS,
\end{equation}
or, after integrating
\begin{equation}
\Phi(x^{\mu}) = \frac{\Lambda^2}{12} \, S^2(x^{\mu}),
\end{equation}
such that $d\Phi(x^{\mu})= -\frac{\Lambda}{2} d\theta(x^{\mu})$. Hence, from eq. (\ref{aac}) we obtain that the quantum volume is given by
\begin{equation}
V_q = \sqrt{-\hat{g}} \, e^{-2\theta} =\sqrt{-\hat{g}} \, e^{\frac{1}{3} \Lambda S^2},
\end{equation}
where $\Lambda S^2 >0$. This means that $V_q \geq \sqrt{-\hat{g}}$, for $S^2\geq 0$, $\Lambda >0$ and $\theta <0$. This implies a signature for the metric: $(-,+,+,+)$ in order for the
cosmological constant to be positive and a signature $(+,-,-,-)$ in order to have $\Lambda \leq 0$.Finally, the action (\ref{aac}) can be rewritten in terms of both, quantum volume and the quantum Lagrangian density ${\cal L}_q = \left[\frac{\hat{R}}{2\kappa} + \hat{{\cal L}}\right]\,e^{2\theta}$
\begin{equation}
{\cal I} = \int d^4 x\, V_q\, {\cal L}_q.
\end{equation}

As was demonstrated in \cite{RB}
the Einstein tensor can be written as
\begin{equation}
\bar{G}_{\alpha\beta} = \hat{G}_{\mu\nu} + \theta_{\alpha ; \beta} + \theta_{\alpha} \theta_{\beta} + \frac{1}{2} \,g_{\alpha\beta}
\left[ \left(\theta^{\mu}\right)_{;\mu} + \theta_{\mu} \theta^{\mu} \right],
\end{equation}
and we can obtain the invariant cosmological constant $\Lambda$
\begin{equation}\label{p}
\Lambda = -\frac{3}{4} \left[ \theta_{\alpha} \theta^{\alpha} + \hat{\Box} \theta\right],
\end{equation}
so that we can define a geometrical Weylian quantum action
${\cal W} = \int d^4 x \, \sqrt{-\hat{g}} \, \Lambda$, such that the dynamics of the geometrical field, after imposing $\delta
W=0$, is described by the Euler-Lagrange equations which take the form
\begin{equation}\label{q}
\hat{\nabla}_{\alpha} \Pi^{\alpha} =0, \qquad {\rm or} \qquad \hat\Box\theta=0,
\end{equation}
where the momentum components are $\Pi^{\alpha}\equiv -{3\over 4} \theta^{\alpha}$ and the relativistic quantum algebra is given by\cite{RB}
\begin{equation}\label{con}
\left[\theta(x),\theta^{\alpha}(y) \right] =- i \Theta^{\alpha}\, \delta^{(4)} (x-y), \qquad \left[\theta(x),\theta_{\alpha}(y) \right] =
i \Theta_{\alpha}\, \delta^{(4)} (x-y),
\end{equation}
with $\Theta^{\alpha} = i \hbar\, \hat{U}^{\alpha}$ and $\Theta^2 = \Theta_{\alpha}
\Theta^{\alpha} = \hbar^2 \hat{U}_{\alpha}\, \hat{U}^{\alpha}$ for the Riemannian components of velocities $\hat{U}^{\alpha}$.

\subsection{Charged geometry and vector field dynamics}

In order to extend the previous study we shall consider that the scalar field $\theta$ is given by
\begin{equation}
\theta(x^{\alpha}) = \phi(x^{\alpha}) \, e^{-i \theta(x^{\alpha})}, \qquad {\rm or} \qquad \theta(x^{\alpha}) = \phi^*(x^{\alpha}) \, e^{i \theta(x^{\alpha})},
\end{equation}
where $\phi(x^{\alpha})$ is a complex field and $\phi^*(x^{\alpha})$ its complex conjugate. In this case, since $\theta^{\alpha} = e^{i\theta} \left(\hat\nabla^{\alpha} + i \theta^{\alpha}\right) \phi^* $, the Weylian connections hold
\begin{equation}\label{con}
\Gamma^{\alpha}_{\beta\gamma} =
\left\{ \begin{array}{cc}  \alpha \, \\ \beta \, \gamma  \end{array} \right\} + e^{i\theta} \, g_{\beta\gamma}\,\left(\hat\nabla^{\alpha} + i \,\theta^{\alpha}\right) \phi^* \equiv
\left\{ \begin{array}{cc}  \alpha \, \\ \beta \, \gamma  \end{array} \right\} +  \,g_{\beta\gamma}\, e^{i\theta} \left( D^{\alpha} \phi^*\right) ,
\end{equation}
where we use the notation $D^{\alpha} \phi^*\equiv  \left(\hat\nabla^{\alpha} + i \theta^{\alpha}\right) \phi^*$. The Weylian components of the velocity $u^{\mu}$ and the
Riemannian ones $U^{\mu}$, are related by
\begin{equation}
u^{\mu} = \hat{ U}^{\mu} + e^{i \theta}\left(D^{\mu} \phi^*\right) \left(x_{\epsilon} \hat{ U}^{\epsilon}\right).
\end{equation}
Furthermore, using the fact that
\begin{equation}\label{del}
\delta g_{\alpha\beta} = e^{-i \theta} \left[
\left( \hat\nabla_{\beta} - i \theta_{\beta} \right) \hat{U}_{\alpha} + \left( \hat\nabla_{\alpha} - i \theta_{\alpha} \right) \hat{U}_{\beta} \right]
\phi \,\delta S,
\end{equation}
we can obtain from the constriction $\Lambda \delta g_{\alpha\beta} = g_{\alpha\beta} \delta \Phi$, that
\begin{equation}
\delta \Phi = \frac{\Lambda}{4} g^{\alpha\beta} \, \delta g_{\alpha\beta} ,
\end{equation}
so that, using (\ref{del}), the flux of $A^{\mu}$ across the Gaussian-like hypersurface can be expressed in terms of the quantum derivative of the complex field:
\begin{equation}\label{flux}
\frac{\delta \Phi}{\delta S} \equiv \frac{d\Phi}{dS} = \frac{\Lambda}{2} e^{i\theta} \hat{U}_{\alpha} \left( D^{\alpha} \phi^* \right).
\end{equation}
Using the fact that $\hat\nabla_{\alpha} \delta W^{\alpha} = \delta\Phi$, it is easy to obtain
\begin{equation}\label{fl}
\hat\nabla_{\mu} A^{\mu} = \frac{\Lambda}{2} e^{i \theta} \hat{U}_{\alpha} \left( D^{\alpha} \phi^* \right),
\end{equation}
where we have defined $A^{\mu}= \frac{\delta W^{\mu}}{\delta S}$. Notice that the velocity components $\hat{U}^{\alpha}$ of the Riemannian observer define the gauge of the system.
Furthermore, due to the fact that $\delta W^{\alpha} = g^{\beta\gamma} \hat\nabla^{\alpha} \delta\Psi_{\beta\gamma}$, hence we obtain that
\begin{equation}\label{gw}
\frac{\delta W^{\alpha}}{\delta S} \equiv A^{\alpha} = g^{\beta\gamma} \hat\nabla^{\alpha} \chi_{\beta\gamma} \equiv \hat\nabla^{\alpha} \chi,
\end{equation}
where $\chi_{\beta\gamma}$ are the components of the gravitational waves:
\begin{equation}
\hat\nabla_{\alpha} A^{\alpha} = g^{\beta\gamma} \hat\nabla_{\alpha} \hat\nabla^{\alpha} \chi_{\beta\gamma} \equiv \hat\Box \chi.
\end{equation}

\section{Quantum field dynamics}

In this section we shall study the dynamics of charged and vector fields, in order to obtain their dynamical equations.

\subsection{Dynamics of the complex fields}

The cosmological constant (\ref{p}) can be rewritten in terms of $\phi=\theta \, e^{i \theta}$ and $\phi^*=\theta e^{-i \theta}$
\begin{equation}\label{cc}
\Lambda = - \frac{3}{4} \left[ \left( \hat\nabla_{\nu} \phi \right) \left( \hat\nabla^{\nu} \phi^*\right) + \theta_{\nu}\, J^{\nu} - \frac{4}{3} \Lambda \phi \phi^*\right],
\end{equation}
where the current due to the charged fields is
\begin{equation}\label{co}
J^{\nu} = i\, \left[\delta^{\nu}_{\epsilon} \left(\hat\nabla^{\epsilon} \phi \right) \phi^* - \phi \left(\hat\nabla^{\nu} \phi^* \right) \right].
\end{equation}
As can be demonstrated, $\hat\nabla_{\nu} J^{\nu} =0$, so that we obtain the condition
\begin{equation}
\phi^* \,e^{i\,\left(\theta-\frac{\pi}{2}\right)} = \phi \,e^{-i\,\left(\theta-\frac{\pi}{2}\right)}.
\end{equation}
The components of the current also can be written in terms of the quantum derivative
\begin{equation}
J^{\mu} = -2\left(1+i\right) \phi \,e^{i\,\theta} \left( D^{\mu} \phi^*\right) \phi,
\end{equation}
where the density of electric charge is given by $J^0$, and the charge is
\begin{equation}
Q = -2\left(1+i\right) \,\int d^3 x \sqrt{|{\rm det}[g_{ij}]|} \,\phi(x^{\alpha}) \,e^{i\,\theta} \left[D^{\mu} \phi^*(x^{\alpha})\right] \phi(x^{\alpha}).
\end{equation}

The second equation in (\ref{q}) results in two different equations
\begin{eqnarray}
&& \left( \hat\Box +i \theta_{\mu} \hat\nabla^{\mu} + \frac{4}{3} \Lambda \right) \phi^* =0, \\
&& \left( \hat\Box - i \theta^{\mu} \hat\nabla_{\mu} + \frac{4}{3} \Lambda \right) \phi =0,
\end{eqnarray}
where the gauge equations are
\begin{eqnarray}
-\left[i \theta_{\mu} \hat\nabla^{\mu} + \frac{3}{4} \Lambda \right]\phi^* & = & \frac{3}{4} \Lambda  \, e^{-i \left(\theta- \frac{\pi}{2}\right)} , \\
\left[ i \theta^{\mu} \hat\nabla_{\mu}- \frac{3}{4} \Lambda \right]\phi & = &  \frac{3}{4} \Lambda  \, e^{i \left(\theta- \frac{\pi}{2}\right)},
\end{eqnarray}
so that finally we obtain the equations of motion for both fields
\begin{eqnarray}
&& \hat\Box  \phi^* = \frac{3}{4} \Lambda \, e^{-i \left(\theta-\frac{\pi}{2}\right)}, \\
&& \hat\Box \phi = \frac{3}{4} \Lambda \, e^{i \left(\theta-\frac{\pi}{2}\right)}.
\end{eqnarray}
Notice that the functions $e^{\pm i \left(\theta- \frac{\pi}{2}\right)}$ are invariant under $\theta = 2 \, n \pi$ ($n$- integer) rotations, so that the complex fields
are vector fields of spin $1$.
Using the expressions (\ref{con}) to find the commutators for the complex fields, we obtain that
\begin{equation}
\left[ \phi^*(x), D^{\mu} \phi^*(y)\right] = \frac{4}{3} i \Theta^{\mu} \,\delta^{(4)}(x-y), \qquad
\left[ \phi(x), D_{\mu} \phi(y)\right] = -\frac{4}{3} i \Theta_{\mu} \,\delta^{(4)}(x-y),
\end{equation}
where $D^{\mu} \phi^{*}\equiv \left(\hat\nabla^{\mu} + i\,\theta^{\mu}\right) \phi^{*}$ and $D_{\mu} \phi\equiv \left(\hat\nabla_{\mu} - i\,\theta_{\mu}\right) \phi$.

\subsection{Dynamics of the vector fields}

On the other hand, if we define $F^{\mu\nu} \equiv \hat\nabla^{\mu} A^{\nu} - \hat\nabla^{\nu} A^{\mu}$,  such that $A^{\alpha}$ is given by (\ref{gw}), we obtain the equations of motion
for the components of the electromagnetic potentials $A^{\nu}$: $\hat\nabla_{\mu} F^{\mu\nu}=J^{\nu}$
\begin{equation}\label{mx}
\hat\Box A^{\nu} -  \hat\nabla^{\nu} \left(\hat\nabla_{\mu} A^{\mu} \right) = J^{\nu},
\end{equation}
where $J^{\nu}$ being given by the expression (\ref{co}) and from eq. (\ref{gau}) we obtain that $\hat\nabla_{\mu} A^{\mu} =-\frac{\Lambda}{2} \theta_{\mu} \hat{U}^{\mu}=\frac{2}{3} \Lambda^2 \, S(x^{\mu})= 4\frac{d\Phi}{dS}$ determines the gauge that depends on the Riemannian frame adopted by the relativistic observer. Notice that for massless particles the Lorentz gauge is fulfilled, but it does not work for massive particles, where $S\neq 0$.

\section{Final remarks}

We have studied charged and electromagnetic fields from relativistic quantum geometry. In this formalism
the Einstein tensor complies with gauge-invariant transformations studied in a previous work\cite{rb}. The quantum dynamics of the fields
is described on a Weylian manifold which comes from a geometric extension of the Riemannian manifold, on which is defined the classical geometrical background.
The connection that describes the Weylian manifold is given in eq. (\ref{con}) in terms of the quantum derivative of the complex vector field with a Lagrangian density described by the cosmological constant (\ref{cc}). We have demonstrated that
vector fields $A^{\mu}$ describe an extended Maxwell dynamics [see eq. (\ref{mx})], where the source is provided by the charged fields current density $J^{\mu}$, with tetra-divergence null. Furthermore, the gauge of $A^{\mu}$ is determined by the relativistic observer: $\hat\nabla_{\mu} A^{\mu} =\frac{\Lambda}{2} \theta_{\mu} \hat U^{\mu}$.

\end{document}